# Directional Spin-wave Propagation in the Skyrmion Chain

Chunlei Zhang,[1, a)] Chendong Jin,[1, a)] Jinshuai Wang,[1] Haiyan Xia,[1] Jianing Wang,[1] Jianbo Wang,[1, 2] and Qingfang Liu[1, b)]

[1]*Key Laboratory for Magnetism and Magnetic Materials of the Ministry of Education, Lanzhou University, Lanzhou 730000, People's Republic of China*

[2]*Key Laboratory for Special Function Materials and Structural Design of the Ministry of the Education, Lanzhou University, Lanzhou 730000, People's Republic of China*

The Dzyaloshinskii-Moriya interaction (DMI), favoring a chiral spin structure like the skyrmion, gives rise to the nonreciprocal propagation of spin waves. We investigate the propagation of spin waves in a nanostripe with the presence of a skyrmion chain by using micromagnetic simulations. Through applying a microwave locally, it is found that when the interval between skyrmions is large enough, the spin waves can be separated to the counter direction according to different frequencies. While for the tightly arranged skyrmions, the skyrmion chain with strong interactions between skyrmions becomes a channel for spin waves, which is around the frequency of skyrmion breathing

---

a) These authors contributed equally to this work.
b) Electronic mail: liuqf@lzu.edu.cn

and exhibit a characteristic of directional propagation. This work opens a vista for skyrmion-based spin wave devices.

## 1. Introduction

The Joule heating originating from the electric current is an insurmountable obstacle in the silicon-based technologies. Studies of spin waves attract considerable attention because the spin waves are free of dissipation from Joule heating[1]. Spin waves are disturbances propagating in the ordering magnetic material and is first predicted by F. Bloch in 1929[2]. The spin waves are also called the magnons, which are the quanta of magnetic excitations in the spin ordering. The magnon existing in the insulating magnetic materials provides a road to wave-based computing technologies[1,3,4], and access to a higher frequency range (from GHz to THz)[5-8].

The Dzyaloshinskii-Moriya interaction (DMI) is an antisymmetric interaction induced by spin-orbit coupling[9,10]. Nowadays, it has been demonstrated that the DMI is widely presented in the non-centrosymmetric magnetic compounds and the ultrathin film with inversion symmetry breaking[11,12]. The antisymmetric interaction favors a noncollinear magnetization and gives rise to the chiral magnetic orders like the skyrmion[13-19], a topological chiral spin texture. The DMI has also a great impact on the dynamical fluctuation of the spin system[20-26]. It is measured by the Brillouin light scattering that the DMI causes the nonreciprocal propagation of the spin waves[27,28]. The theoretical and experimental works in recent years also show that the DMI brings nontrivial topologies of the magnons, which exhibits the unidirectional and stable characteristics in the bulk edge[29-32]. The type of magnons are called the topological magnons and is promising to the future spintronics applications[33].

Regardless of the magnons with the topological property, earlier literatures have revealed that the presence of DMI causes the antisymmetry of the magnon propagating in the counter-propagation directions in the films[23,24,26,34,35]. The fundamental mechanism attracted extensive interests and attributes to ingenious designs for the spin wave devices[34-36]. The aim of this work is to explore a new way to achieve the manipulation of spin waves by introducing skyrmions. The work examined the spin waves in a nanostripe with linearly arranged skyrmions by numerically solving the Landau-Lifshitz-Gilbert (LLG) equation. Applying a local microwave in the nanostripe with the skyrmion chain, the spin waves exhibit some interesting phenomenon by adjusting the interval between skyrmions, due to the modification and magnetization oscillation of skyrmions[35,37-41].

## 2. Simulation model

We take the schematic structure shown in Fig. 1 as a simulation model. A linear array of equal-interval skyrmions is arranged in a nanostripe, which is a perpendicularly magnetized Co ultrathin film with the presence of the interfacial DMI. The interval $d$ between skyrmions only refers to the transverse distance between the centers of nearest neighbor skyrmions. The skyrmion chain spontaneously stabilizes to the center of the stripe. The micromagnetic simulations are performed by using Object Oriented MicroMagnetic Framework (OOMMF) code[42]. The software numerically solves the Landau-Lifshitz-Gilbert equation of motion for the magnetization dynamics

$$\frac{d\bm{m}}{dt} = -\gamma \bm{m} \times \bm{H}_{eff} + \alpha(\bm{m} \times \frac{d\bm{m}}{dt})$$

where $\bm{m}$ is the unit vector along the local magnetization, $\gamma$ is the gyromagnetic ratio, $\alpha$ is the Gilbert damping constant, and $\bm{H}_{eff}$ is the local effective magnetic field including the exchange field, anisotropy field, magnetostatic field, Zeeman field, and DMI field. We adopt $A$ = 15 pJ/m as the exchange constant, $D$ = 3.0 mJ/m$^2$ as DMI constant, $K$ = 800 kJ/m$^3$ for perpendicular magnetic anisotropy, $M_s$ = 580 kA/m as saturated magnetization and the coefficient $\alpha$ = 0.01 for Gilbert damping. For the simulation, the whole system is divided into 2100×20×1 unit meshes, which corresponds to a mesh size of 2×3×0.6 nm³ within the range of the Bloch exchange length $\lambda = \sqrt{A/K} = 4.33$ nm [43]. The origin of the Cartesian coordinate frame is set at the center of the stripe. For exciting spin waves, an excitation pulse, $h_y = h_0\mathrm{sinc}(2\pi ft)$, with maximum amplitude $h_0$ = 0.5 mT and cut-off frequency $f$ = 100 GHz, was applied locally to the central section of the stripe[44,45].

## 3. Results and discussion

In the case that a skyrmion in the 60 nm wide square, the spin waves are generally composed by two parts, i.e., excitation of the ferromagnetic (FM) region and the skyrmion, as shown in Fig. 2. The microwave-absorption peak at the 16.9 GHz denotes the skyrmion breathing mode and the two higher-frequency peaks are the resonance modes of FM region. The situation is similar in our model shown in Fig. 1, whose skyrmion breathing mode is lower than 20 GHz and far below the spin-wave excitation frequency of FM material[35]. And therefore the spin waves of the two parts in the nanostripe are completely detached.

### 3.1 Spin-wave separation in the high frequency band

To obtain the spin wave dispersion relations, we took a two-dimensional Fourier transform of $\delta m_z$ for the nanostripe with the interval $d$ = 60 nm between skyrmions. Figures 3(a) - (i) present an overview of dispersion relations at the different positions from $y$ = 30 to –30 nm. As for the present model, the skyrmion breathing mode under 20 GHz is extremely weak in the dispersion curves and it will be discussed in the 3.2 section. Therefore, the magnetization fluctuation of whole stripe is mainly from the FM region[35]. In contrast to the nanostripe without skyrmion in the FM state (supplementary material), the dispersion curves are observed feature bandgaps. These band gaps reveal that the stripe with skyrmion chain is an artificial magnonic crystal, and the skyrmions work as periodic potential wells for magnons[35].

To observe the dispersion relations in detail, we use different color bar for each figure. From Figs. 3(a) and (d), we can see several bandgaps and the spin waves higher than 76.4 GHz shows a characteristic of unidirectional propagation at the edges of the stripe. The discrete frequency bands shown in the Figs. 3(b), (c), (e) and (f) also retain their own directions. When it comes to the interior of the stripe, for Figs. 3(b), (c), (e), (f), (g) and (i), the dispersion curves show repetitive patterns in the neighboring Brillouin zones. The crossed and repetitive curves in each figures suggest that the stripe with skyrmion chain is resemble to the helical spin structure[40]. Comparing the Figs. 3(a) - (c) and Figs. 3(d) - (f), we divide the dispersion relations into the two groups, $y > 0$ and $y < 0$, then the dispersion relations of the two sections are completely reversed. The nonreciprocality of the spin waves is owing to the existence of the DMI. The spatial symmetry of dispersion relations leads to a speculation: ignoring the spatial distribution of the spin waves in the nanostripe, we can obtain the same signal when spin wave detectors are put in the two ends (left and right).

While for the case $y = 0$ nm in Fig. 3(h), the curves show an apparent dispersion relation with two modes of spin wave. In fact, the dispersion relation in the center of nanostripe should be a symmetrical curve in theory, as shown in ref. [35]. It should be noted that our model is discretized in 2100×20×1 meshes, therefore we cannot attain the dispersion relation in the center ($y = 0$ nm) directly and the relation is an average of the figures $y = 3$ and $–3$ nm. This average is only a result of simple mathematical means and cannot avoid from deviations. The spin waves fitting in the red dot line in Fig. 3(h) is the 2nd-order spin waves, which is in the range 72.8 - 75.9 GHz. Compared with the band in the red dot line in Figs. 3(c) and (f), they share the similar shape and frequency which is ranging from 72.8 - 75.9 GHz. However, the spin waves marked by red dot line in Figs. 3(c) and f show a direction of propagation opposite to the spin waves below 69.6 GHz, which can cause a separation of spin waves in different frequencies.

Figure 4 describes the dependence of intensity and frequency for specified directions $k_{-x}$ and $k_x$ in the whole stripe. Different from the detector in above, it provides the spatial distribution of spin waves and the spin waves for the two directions are fully symmetry. It shows that the spin wave is intensive in the center and edge of the stripe, as the skyrmions and edges are easier to be excited. There still exist some spin waves in the region of 10 nm < | y | < 30 nm in spite of their much lower intensity (that shown in color bars in Figs.3(b) and (e)).

In Fig. 4, the spin waves in the center of nanostripe are generally divided into five parts by frequency. Respectively, the five parts are 72.8 GHz < $f$ < 75.9 GHz, 69.6 GHz < $f$ < 70.9 GHz, 64.2 GHz < $f$ < 68.1 GHz, 61.0 GHz < $f$ < 63.3 GHz and 49.5 GHz < $f$ < 53.3 GHz. For each figure in Fig. 4, the part ranging from 72.8 to 75.9 GHz is spatially separated versus to the other frequency. That means if a spatial bias pulse is applied to the region, –2nm < $x$ < 2nm and –30nm ≤ $y$ < 0, the spin waves ranging from 72.8 to 75.9 GHz will propagate along the $k_x$ direction, which is counter to the spin waves lower than 72.8 GHz.

The above result has demonstrated that the spin-wave separation is achievable, but the origin of the separated spin waves with frequency ranging from 72.8 to 75.9 GHz is still unknown. According to dispersion curves in Fig. 3(h), they correspond to the 2nd-order dispersion relation and share a similar form in Figs. 3(c) and (f) (marked by the red dot line). Therefore, we speculate that these separated bands of spin waves are detached from the 2nd-order spin waves. We confirm the speculation in the following paragraphs.

While for the same stripe in the FM state with DMI, spin waves of the same frequency tend to propagate reversely for the different orders (detailed information is included in the supplementary material). The dispersion relation of nanostripe with the skyrmion chain is the modulation of the original dispersion relation in the FM state. Therefore, when the unmodulated dispersion relation of the order 2 is changing with the width of FM stripe, the spin waves in the stripe with skyrmion chain shall show a similar tendency. The dependence of the separated frequency band on width is shown in Fig. 5(a) and the spin wave with the highest intensity among the separated band is selected as the featured frequency of the band. Figure 5(a) shows that the fundamental frequency of the 1st-order (mode n = 1) spin wave is nearly a constant, both the 2nd-order (mode n = 2) fundamental frequency and the separated frequency decrease as the stripe width increasing. In addition, we also observed that the strength of separated frequency decreases with a narrower stripe, as shown in Fig. 5(b). The result is in agreement with the previous observation[35] that the 2nd-order dispersion curve was not observed in the skyrmion chain with skyrmion interval $d$ = 40 nm. Overall, these results ensure that the separated frequencies in Fig. 4 are originated from the spin waves of the 2nd-order. Finally, we can make a reasonable inference for the origin of frequencies separation: with the modulation of equally spaced skyrmions, parts of the 2nd-order spin waves are allowed in the stripe when the 1st-order spin waves with the same frequencies are prohibited as the width of nanostripe is over 48 nm.

**3.2 Directional propagation of spin wave in the low frequency band**

We have mentioned that skyrmions can also participate in the dynamic excitations. The mechanism of the dynamic skyrmion excitations have been studied both in experimental and theory [37-39,41]. They imply that the skyrmions can also be regarded as the magnon carrier and the coupled skyrmions can also be a natural transmission channel of magnons[46,47]. Based on those works, we decrease the interval between two skyrmions to form a magnonic channel. The channel is peculiarly for the skyrmion excitation whose frequency is much lower than the spin-wave frequency in the skyrmion based magnonic crystal.

With the decreasing interval between skyrmions, the skyrmions chain is misaligned by the interaction between closely arranged skyrmions, as shown in Fig. 7(a). As for a single skyrmion, there is an intensive microwave-absorption peak at the 16. 9 GHz corresponding to the breathing mode of skyrmion (shown in Fig. 2). Figure 6 exhibits the dependence of intensity and frequency for specified directions $k_{-x}$ in the whole stripe. According to Fig. 6, an intensive response ranging from 12.1 to 18.9 GHz is found in the stripe when the skyrmion interval $d$ was fixed to 30 nm. So the excitation on the stripe is surely the breathing mode of skyrmions (more information in supplementary material). Compared with the spatially localized spin wave with high frequency in the Fig. 6, the dynamic excitation of skyrmion in the region $-10$ nm $< y < 0$ nm and $0$ nm $< y < 10$ nm is almost symmetric, as skyrmion excitation is the magnetization oscillation of the entire spin texture.

When the microwave field with the frequency ranging from 12.1 to 18.9 GHz was applied to the region, $-2$ nm $< x < 2$ nm and $-30$ nm $\leq y < -3$ nm (Fig. 7(a)), the excited spin waves exhibit a characteristic of directional propagation. By a two-dimensional Fourier transform of the magnetization component $m_z$ in space and time, we can visualize the spatial distribution of the dynamic magnetization component $m_z$, as shown in the left part of Fig. 7(b). The right part of Fig. 7(b), correspond to the left respectively, is the spatial distribution of $m_z$ in the $x$ axis when $y = 3$ nm and denoted the misalign of skyrmions. The data in Fig. 7(b) reveals that spin waves propagate directionally along the –*x* for the interval *d* < 36 nm. While the misalign of skyrmion is disappeared with the interval *d* = 36 nm, the spin waves are prohibited.

A qualitative model can be used to explain the directional propagation of spin waves on the skyrmion chain, as shown in Fig. 8. Figure 8(a) is a propagation unit in the spin wave channel. When the interval between skyrmions increases to 36 nm, the transmission on skyrmion chain stops instantaneously, as shown in Fig. 7(b). And with the interval increasing, the misalignment between skyrmions is also gradually disappeared, which denotes the decline of the interaction between skyrmions. In Fig. 8(a), it shows that the skyrmion interval is large enough and the interaction between skyrmion is weak, so the excited spin waves cannot propagate through the interaction between skyrmions, as the spin waves ranging from 12.1 to 18.9 GHz is banned in the regions of the FM state between the skyrmions. The explain agrees well with the prohibition of spin waves in Fig. 7(b) and suggests that the spin wave channel only exists when the coupling between skyrmions is large enough.

Figure 8(b) is the excitation unit while the interval is small. The microwave field is applied to the region betwixt the skyrmions. The magnetization in this area can be described by a one-dimensional schematic structure as shown in the orange dotted square in Fig. 8(b). Actually, the structure in the region is resemble to a chiral domain wall and it performs like a spin-wave diode[34]. According to the work of the spin-wave diode, the domain wall with DMI shall cause the spatial separation of the spin-wave bound states and the chiral domain wall like structure in our work plays the same role. With the microwave field $h_y$ (applied in the region –2 nm < *x* < 2 nm and –30 nm ≤ *y* < –3 nm indicated by the blue square in Fig. 8(b)), the initially excited spin waves will be separated by the domain-like structure and propagate along the *y* axis in opposite directions. The region of the applied microwave field is a high-energy area filled with magnons. The spin waves are more likely to propagate out from the excitation source ($k_y$) and the spin wave with $k_{-y}$ whose direction points to the excitation source is suppressed. Owing to the reason, the excitation source is similar to an energy barrier for the longitudinal spin wave propagating through the microwave field. As the one-dimensional schematic structure is composed of the skyrmion edges, the antiparallel spin waves with different intensity at the two edges lead to different intensity of skyrmion breathing near the excitation source.

Figures 9(a) and (b) provide two pieces of evidence for our explanations. Since the initially excited spin waves are transmitted in longitude, the suppressed spin wave can be relieved by reducing the longitudinal length $c$ of the energy barrier (excitation source). As shown in Fig. 9(a), the relative amplitude (the ratio of maximum amplitude between spin waves propagating to left and right) decreases as the longitudinal length $c$ of the excitation source decreasing. Moving the excitation source transversely, Fig. 9(b) shows the spin waves excited at the ends of the skyrmion chain. Fig.9(b) shows the $\delta m_z$ of both ends of the skyrmion chain when the excitation source is moved transversely to the end of nanostripe. The maximum amplitude of the right is about twice of the left. The phenomenon shows that the chiral domain like structure surely suppresses the spin waves propagating to the left and allow the spin waves to the right. These confirm our qualitative explanation about the directional propagation of spin waves.

## 4. Conclusion

In summary, we investigate the directional spin-wave propagation in the skyrmion chain by micromagnetic simulations. The skyrmion has its own excitation mode and the frequency of the mode is much lower than excitation of FM state. For the high frequency band with the large interval, the nanostripe work as a magnonic crystal. We verified that the skyrmion chain can split continuous dispersion curves and separate the 2nd-order spin wave from the 1st-order spin wave. Since the spin waves of different orders propagate in the counter-propagation directions, the spin waves are separated according to the frequency. With respect to the tightly arranged skyrmion chain, the strongly coupled skyrmions are the propagation channel for the skyrmion breathing mode. Through the spatial separation of the spin-wave bound states caused by the DMI in the non-uniform magnetic structure between skyrmions, the spin waves propagate directionally in the skyrmion chain. Our findings open a new door for skyrmion-based spin-wave devices.

**Acknowledgements**

We would like thank Yuqing Li for the suggestion of part 3.2. This work was supported by National Natural Science Fund of China (11574121 and 51771086).

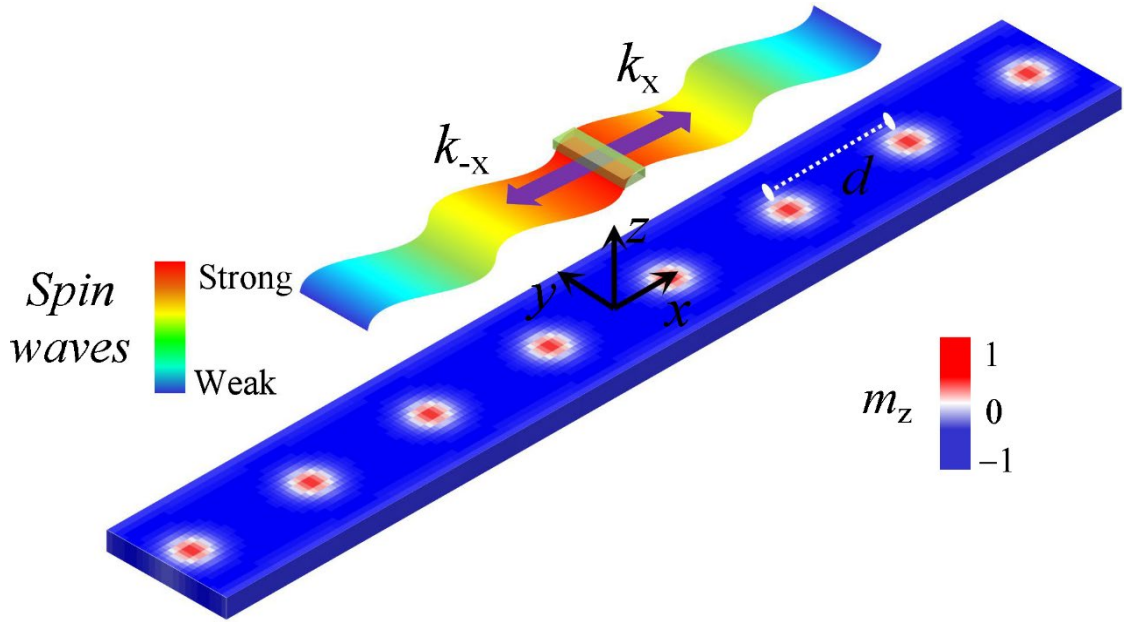

**Fig. 1.** Skyrmions are equally spaced arranged in the center of a 4200×60×0.6 nm³ thin film stripe with perpendicular magnetic anisotropy. The interval between adjacent skyrmions is set to a constant $d$ = 60 nm. The origin of the Cartesian coordinate frame is set at the center of the stripe. The light blue cube represents the applied in-plane magnetic pulse.

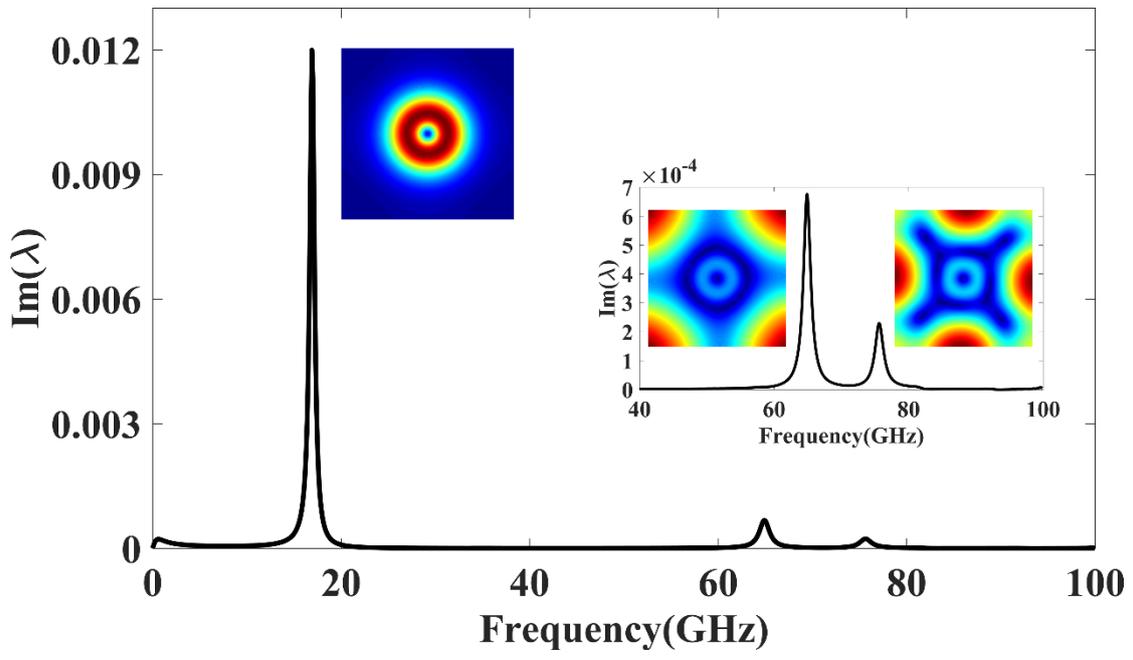

**Fig. 2.** The magnetic spectrum of a skyrmion confined in a 60 nm wide square, which shows microwave-absorption peaks in the frequencies of 16.9, 64.9 and 75.6 GHz. The three insets beside the peaks are the absolute value of δ*m* under the corresponding microwave.

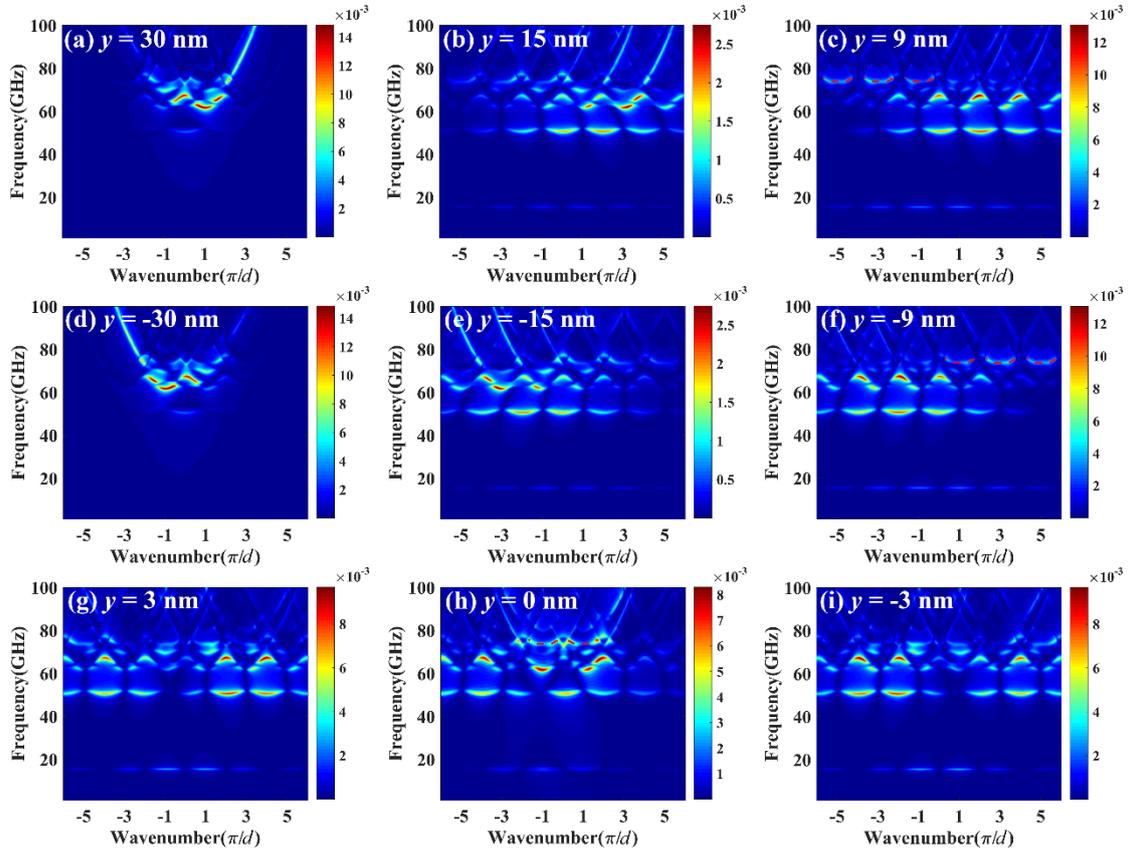

**Fig. 3.** (a)-(i) The dispersion relations of spin waves at the different position in the model proposed by Fig. 1. The interval between adjacent skyrmions is fixed to a constant $d = 60$ nm.

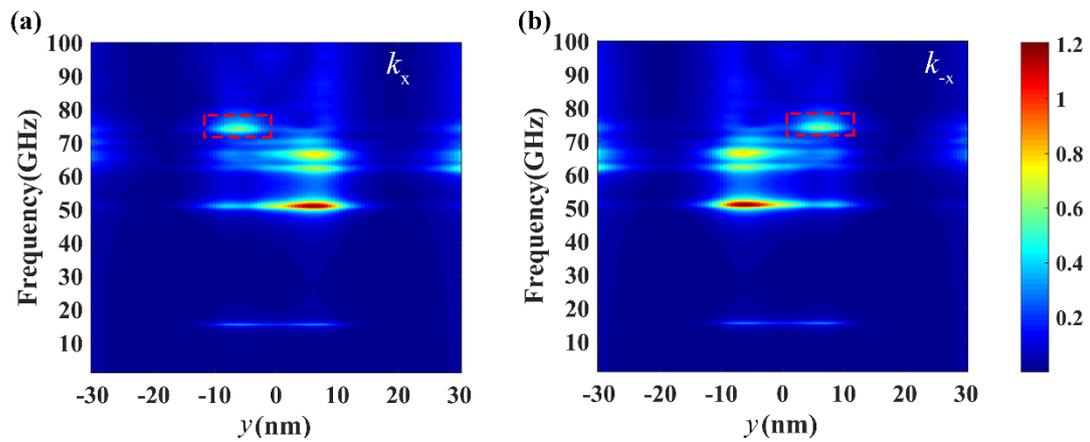

**Fig. 4.** With the skyrmion interval $d$ fixed to 60 nm, the dependence of the spinwave frequency as a function of position as the wavenumber is $k_x$ (a) and $k_{-x}$ (b). The frequency range from 72.8 to 75.9 GHz is noted by the red dotted square.

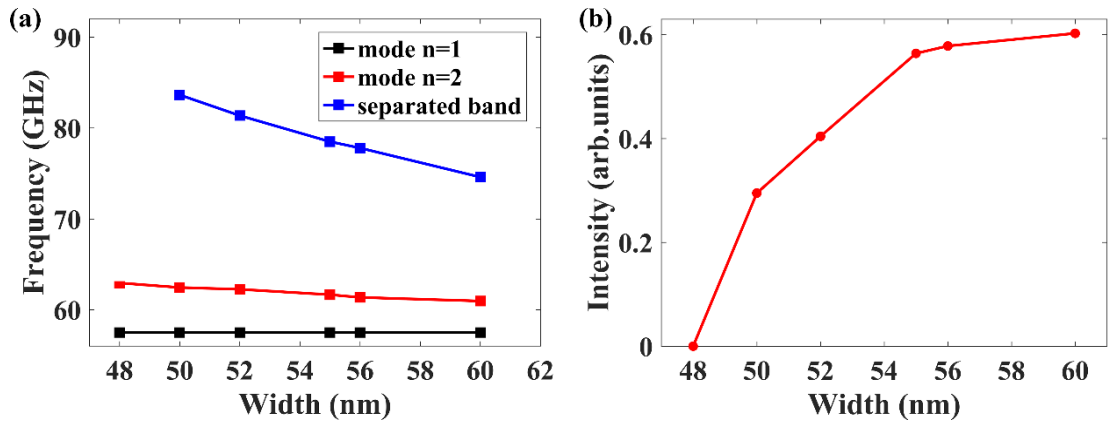

**Fig. 5.** (a) The relation between the three feature frequencies and the stripe width. The feature frequency of the separated band is the most intensive spin wave among the frequency from 72.8 to 75.9 GHz. The mode 1 and 2 is the fundamental frequency of the 1st-order and 2nd-order spin waves respectively. (b) The intensity of the separated band versus the stripe width.

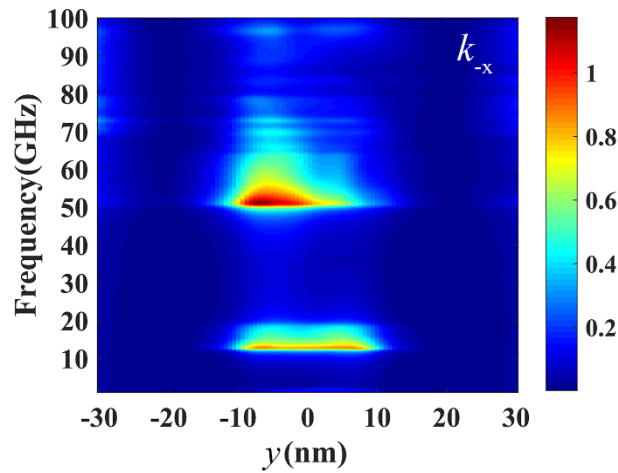

**Fig. 6.** The relationship between the frequency and the position of the $k_{-x}$ spin wave and the interval $d$ of adjacent skyrmions is fixed to 30 nm.

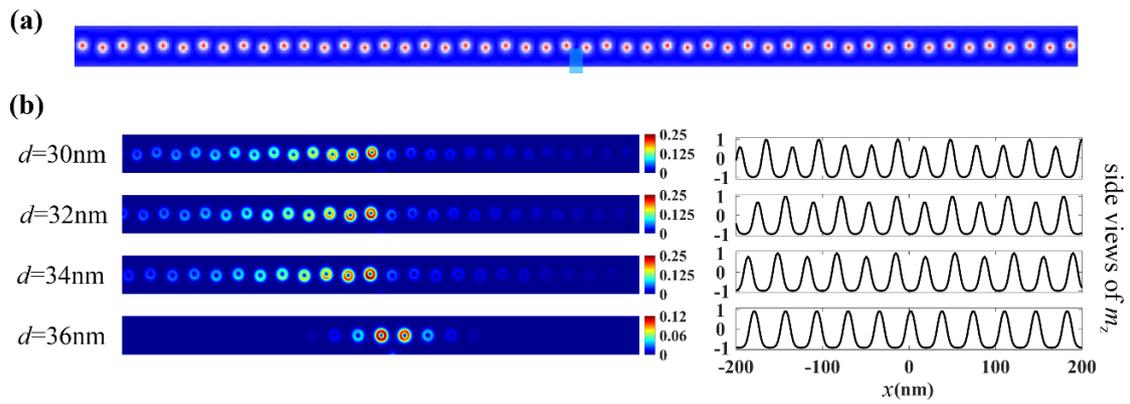

**Fig. 7.** (a) The model of skyrmion chain when the skyrmion interval in transverse is decreased to 30 nm. A microwave field in the region, –2 nm < $x$ < 2 nm and –30 nm ≤ $y$ < –3 nm, is applied on the nanostripe and shown by the transparent blue square. (b) Under the field shown in the figure (a), the absolute value of δ$m_z$ for different skyrmion intervals. The right part, the side views of the left parts, shows the $m_z$ at $y$ = 3 nm from $x$ = –200 to 200 nm.

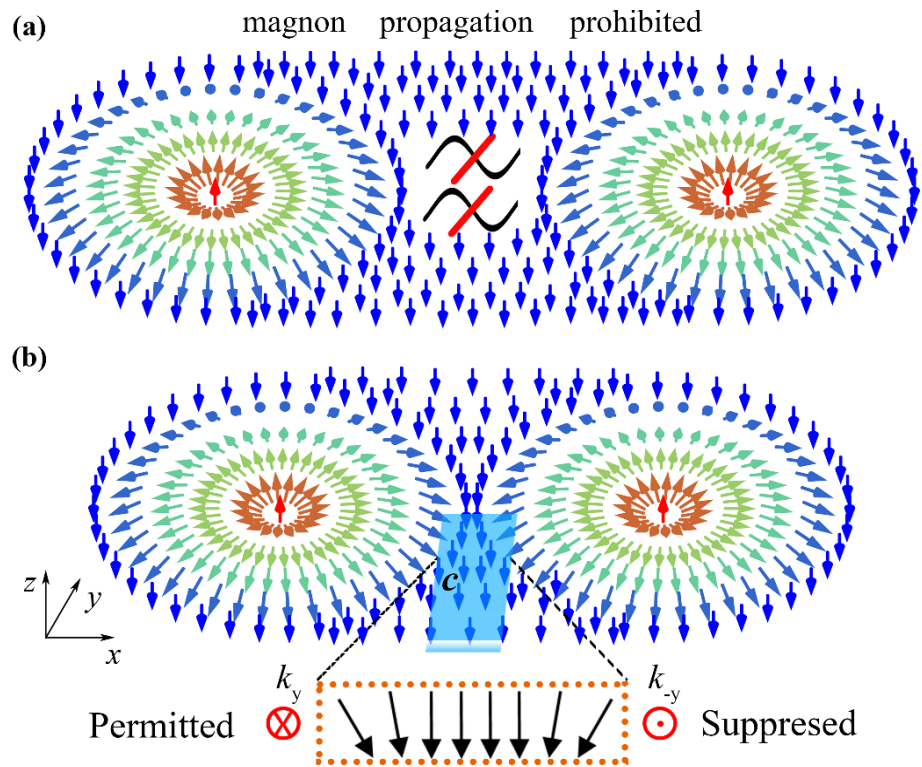

**Fig. 8.** The schematic of the excitation and propagation units of skyrmion chain. (a) The propagation unit while the interval is large enough. There is nearly no interaction between skyrmions and the spin waves cannot propagate by the interaction between skyrmions. (b) The excitation unit while the interval is diminutive. The transparent blue cube represents the excitation area. The orange dotted square shows a one-dimensional schematic of spatial magnetization distribution between skyrmions. The excited spin waves are separated into two directions and the spin wave whose direction points to the excitation source is suppressed. The longitudinal length of the exciting source is marked as $c$.

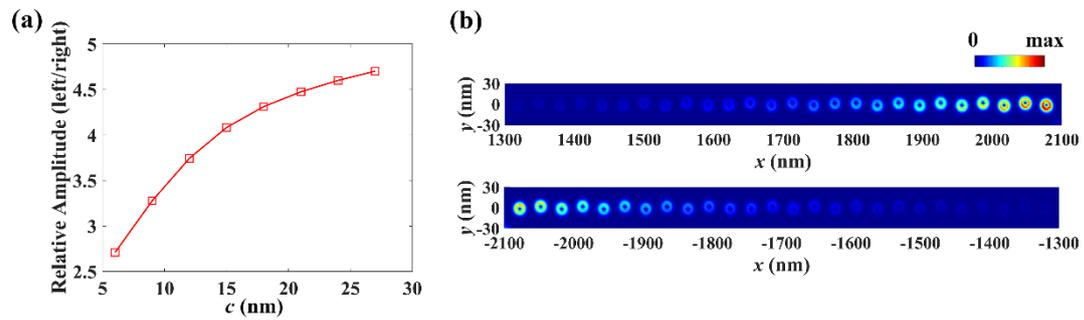

**Fig. 9.** (a) The dependence of the relative amplitude of left to right as a function of the longitudinal length $c$ (shown in Fig. 6(b)). (b) Moving the excitation field transversely to the ends of the skyrmion chain, the absolute value of $\delta m_z$ for the right end (upper of (b)) and the left end (bottom of (b)), and they share the same color bar.